\documentstyle[eqsecnum,aps,epsf]{revtex}
\begin{document}
\newcommand\lsim{\mathrel{\rlap{\lower3pt\hbox{\hskip1pt$\sim$}}
    \raise1pt\hbox{$<$}}}
\newcommand\gsim{\mathrel{\rlap{\lower3pt\hbox{\hskip1pt$\sim$}}
    \raise1pt\hbox{$>$}}}

\title{Limits on Cosmic Chiral Vortons}
\author{C. J. A. P. Martins\thanks{Also at C. A. U. P.,
Rua do Campo Alegre 823, 4150 Porto, Portugal.
Electronic address: C.J.A.P.Martins\,@\,damtp.cam.ac.uk}
and
E. P. S. Shellard\thanks{Electronic address:
E.P.S.Shellard\,@\,damtp.cam.ac.uk}}
\address{Department of Applied Mathematics and
Theoretical Physics\\
University of Cambridge\\
Silver Street, Cambridge CB3 9EW, U.K.}
\maketitle

\begin{abstract}
We study chiral vorton production for
Witten-type superconducting string models in the context of a recently
developed analytic formalism. We delineate three distinct scenarios:
First, a low energy regime (including the electroweak scale) where vortons
can be a source of dark matter. Secondly, an intermediate energy regime 
where the vorton density is too high to be compatible with the standard
cosmology (thereby excluding these models). Finally, a high energy regime
(including the GUT scale) in which no vortons are expected to form. 
The vorton density is most sensitive to the order of the
string-forming phase transition and relatively insensitive to the 
current-forming transition. For a second-order string transition, vorton 
production is cosmologically disastrous for the range $10^{-28}\lsim G\mu
\lsim 10^{-10}$ ($10^{5}\hbox{GeV}\lsim T_{\rm c}
\lsim 10^{14}\hbox{GeV}$), while for the first-order case we 
can only exclude  $10^{-20}\lsim G\mu
\lsim 10^{-14}$ ($10^{9}\hbox{GeV}\lsim T_{\rm c}
\lsim 10^{12}\hbox{GeV}$). We provide a fitting formula which summarises our results.
\end{abstract}

\section{Introduction}
\label{v-in}
It is well known that cosmic strings can (and are perhaps likely to) behave as
`superconducting wires' carrying large currents and charges \cite{witten}.
The charge carriers can be either bosons or fermions (see \cite{vs} for a
review). String superconductivity has a number of important cosmological
consequences, the most important of which is the possibility of forming
vortons \cite{vor}. These are stationary loops (stabilised by angular momentum) that do not radiate classically. At large distances they look like point particles with quantised charge and angular momentum. The reason why they are important \cite{vor,vor2,vornew}
comes from the fact that they redshift like matter (as opposed to radiation), 
and they can be a dark matter candidate. However, it is also
possible that so many of them are produced that the universe becomes dominated
by matter much earlier than in the standard scenario (in disagreement with
observations). In this case, this can provide rather strict constraints 
on allowed particle physics models.

In the past few years, a formalism has been developed \cite{ms,ms2,vornew}
which allows one to calculate vorton densities in any chosen superconducting
string model, subject to some approximations---we refer the reader to
\cite{vornew} for a detailed discussion of the formalism.
Very briefly, this is a two stage process. First, one studies the evolution of
a number of relevant loop solutions to the required microscopic 
equations of motion, and from the results of this analysis one
introduces parameters characterising the loop's ability to evolve into a
vorton state. Secondly, one uses a
phenomenological model for the evolution of the superconducting
currents on the long cosmic string network \cite{mss}, to estimate the
currents carried by string loops formed at all relevant times, and thus
decide if these can become vortons and calculate the corresponding density.

In a previous paper \cite{vornew} we have conjectured that GUT-scale
superconducting string networks do not produce any vortons, while a network
formed at the electroweak scale can produce vortons that make up to $6\%$ of
the critical density of the universe. Here we extend this work by analysing
also the intermediate energy scales, and discussing the corresponding
cosmological bounds. Throughout this paper we will use fundamental units in which
$\hbar=c=k_B=Gm^2_{Pl}=1$, and all logarithms will be base ten.

\section{Witten's superconducting model}
\label{v-wt}
The idea that a cosmic string can be superconducting is originally
due to Witten \cite{witten}, who was also the first to derive a low-energy
effective action for them.
One finds that they are describable by the following Lagrangian density
\begin{eqnarray}
{\cal L}=-\mu_0+\frac{1}{2}\phi_{,a}\phi^{,a}
-A_\mu J^\mu\, . \label{witact}
\end{eqnarray}
Here, the three terms are respectively the usual Goto-Nambu term, the inertia of the charge carriers and the current coupling to the electromagnetic potential.
The microscopic string equations of motion can then be obtained by in
the usual (variational) way.
Numerical simulations of contracting string loops at fixed charge and winding number have shown \cite{davsh} that a `chiral' state with equal charge and current densities is approached as the loop contracts. For this reason, we will
restrict ourselves to the chiral case for the remainder of this letter.

In this limit, introducing the simplifying function $\Phi$ defined as
\begin{equation}
\Phi=\frac{{\dot\phi}^2}{\mu_0 a^2(1-{\bf {\dot x}}^2)}\, ,\label{defphi}
\end{equation}
and choosing the standard 'temporal transverse' gauge the string equations of motion in an FRW background with the line element $ds^2=a^2(d\tau^2-{\bf{dx}}^2)$ have the form
\begin{equation}
\left[\epsilon\left(1+\Phi
\right)\right]{\dot{}} +\frac{\epsilon}{\ell_d}{\dot {\bf x}}^2 =
\Phi'-2\frac{{\dot a}}{a}\epsilon\Phi \, , \label{witchirti}
\end{equation}
(with dots and primes respectively denoting derivatives with respect to the
time-like and space-like coordinates on the worldsheet as usual)
and
\begin{equation}
\epsilon\left(1+\Phi\right) {\ddot {\bf x }}+\frac{\epsilon}{\ell_d}
(1-{\dot {\bf x}}^2) {\dot {\bf x}}=\left[\left(1-\Phi\right)
\frac{{\bf x}'}{\epsilon}\right]'+
\left({\dot \Phi}+2\frac{{\dot a}}{a}\Phi\right){\bf x}'
+2 \Phi{\dot{\bf x}}' \, . \label{witchirsp}
\end{equation}
For simplicity we have introduced the `damping length'
\begin{equation}
\frac{1}{\ell_d}=a\left(2H+\frac{1}{\ell_{\rm f}}\right)\, .\label{defdaml}
\end{equation}
Notice that this includes a frictional term due to particle scattering on
strings (see \cite{ms} for a discussion of the inclusion of this term).
As shown in \cite{mss}, plasma effects are subdominant, except possibly in
the presence of background magnetic fields---either of `primordial' origin
or generated (typically by a dynamo mechanism) once proto-galaxies have formed.
Hence one expects Aharonov-Bohm scattering to be the dominant effect.

\section{Calculating vorton densities}
\label{v-den}
The calculation of vorton densities is a two-stage process. This has been
detailed in \cite{vornew}, so in the present letter we will restrict ourselves
to a brief summary of the ideas involved.
The first stage consists of studying the microphysics of the particular model that one is interested in, in order to derive its microscopic equations of motion and to numerically study some of its solutions. From this one can
establish criteria to decide which loop configurations will produce vortons.
Pending a detailed study of the quantum-mechanical stability of these objects, our criterion should be that loops whose velocity is always small will become vortons, while those which are (sometime during their evolution) relativistic will suffer significant charge losses, so that one cannot realistically expect them to become vortons. In this letter we take `relativistic' to mean a velocity larger than $v_{vor}=0.5$ (see \cite{vornew} for a discussion of the effect of varying this cutoff). The outcome of this task is to pick out the region of the tri-dimensional parameter space formed by the epoch of loop formation $t_i$, its initial size $\ell_i$ and its winding number $W$ where vortons can form.

The second stage is to determine the actual properties of the loops produced by the relevant network. The evolution of the string network (including the mechanism of loop production) is described via the velocity-dependent one-scale model \cite{ms,ms2}. This is a quantitative model
which has two key macroscopic (or `thermodynamic') quantities:
the long-string correlation length $\rho_{\infty}\equiv\mu/L^2$ (where $\mu$ is the string mass per unit length) and the string RMS velocity, $v^2\equiv\langle{\dot{\bf x}}^2\rangle$. The evolution equations for these
can be derived from the microscopic string equations of motion \cite{ms,ms2}, at the expense of introducing two `phenomenological' parameters, a `loop chopping
efficiency' ${\tilde c}$ and a `small-scale structure parameter' $k$.

In addition, we use a simple model for the evolution
of the superconducting currents on the long strings \cite{mss,vornew}. Assuming that there
is a `superconducting correlation length', denoted $\xi$, which measures the scale over which one has coherent current and charge densities on the strings, we can define $N_L$ to be the number of uncorrelated current regions per long-string correlation length, that is $N_L=L/\xi$. By considering how the dynamics of the string network affects $N_L$, one can and obtain an evolution equation for it. Again, one needs to introduce a further phenomenological parameter (denoted $f$ in \cite{mss,vornew}), which quantifies the importance of `equilibration' between neighbouring current regions.

Now, the winding number can be written as $2\pi\Sigma=N_L^{1/2}$ (where $\Sigma$ is a model dependent constant) so that the outcomes of these two steps can be easily related: knowing the properties of the loops produced by the string network of interest and having decided which of these will produce vortons, it is a simple matter to determine the corresponding density.
We note that there are some uncertainties in the values of the
`phenomenological' parameters mentioned above, but these do not
significantly affect the numerical results.

Although in our previous work \cite{vornew} we restricted ourselves to
calculating vorton densities for GUT and weak-scale strings, a number of
general points about the vorton problem became apparent. The most important one is that since loops must be (loosely speaking) non-relativistic if they are to
form vortons, the efficiency of vorton production of a given string network will
vary with time. For the purposes of the present discussion, this efficiency can
be defined as the fraction of all the loops produced by a string network in a
given Hubble time that will eventually become vortons. In fact, the efficiency
will be largest just after the string-forming and superconducting phase
transitions (assumed to occur at about the same temperature, since this is the
most favourable case---see \cite{vornew}), and decreases afterwards. This is
because the relative importance of the frictional force on the string dynamics
decreases with time. In particular, no vortons will be formed once the network
has ceased to be friction-dominated.  Since the duration of this
friction-dominated phase increases as the string mass parameter $G\mu$
decreases, we expect that overall GUT-scale networks will be the less efficient
ones, while electroweak-scale networks will be the most efficient. Indeed,
we have shown in \cite{vornew} that no GUT-scale vortons are expected to
form, while most electroweak-scale loops become vortons.

The second general point to be made is that relatively high currents are needed
on the string loops to prevent them from becoming relativistic (and hence liable to charge losses). Indeed, as was first pointed out in \cite{vornew}, unless
there are `outside' sources (such as primordial magnetic fields), the required
currents are much harder to get than was previously expected. Given this need
to have very large currents, it turns out that the order of the string-forming
and superconducting phase transitions is the crucial parameter determining the
efficiency of vorton formation for a string network of a given energy scale.
In particular, this is much more important than any of the `phenomenological'
parameters used in the analytic modelling of these networks \cite{ms,ms2,vornew}. As one could intuitively have guessed, it is comparatively much harder to form vortons if the phase transitions are of first order, since in this case the early stages of evolution of the network occur in a `stretching' regime \cite{ms2} where only a few relatively large loops are produced with quite small currents. Furthermore, the order of the string-forming transition is the most crucial of the two---the effects of the superconducting phase transition are quickly `forgotten' by the string network.

\section{Vorton densities for all energy scales}
\label{v-new}

In this letter, we extend our previous work by performing the calculation of the
vorton density for typical cosmic string networks of all relevant masses, from
the GUT to the electroweak scale. Since the detailed `algorithm' for the 
calculation has been presented elsewhere \cite{vornew} (and summarised above),
we now simply proceed to present and discuss the results. We will be discussing the two `extreme' cases where the string-forming and superconducting phase transitions are both of first or second order.

\begin{figure}
\vbox{\centerline{
\epsfxsize=0.7\hsize\epsfbox{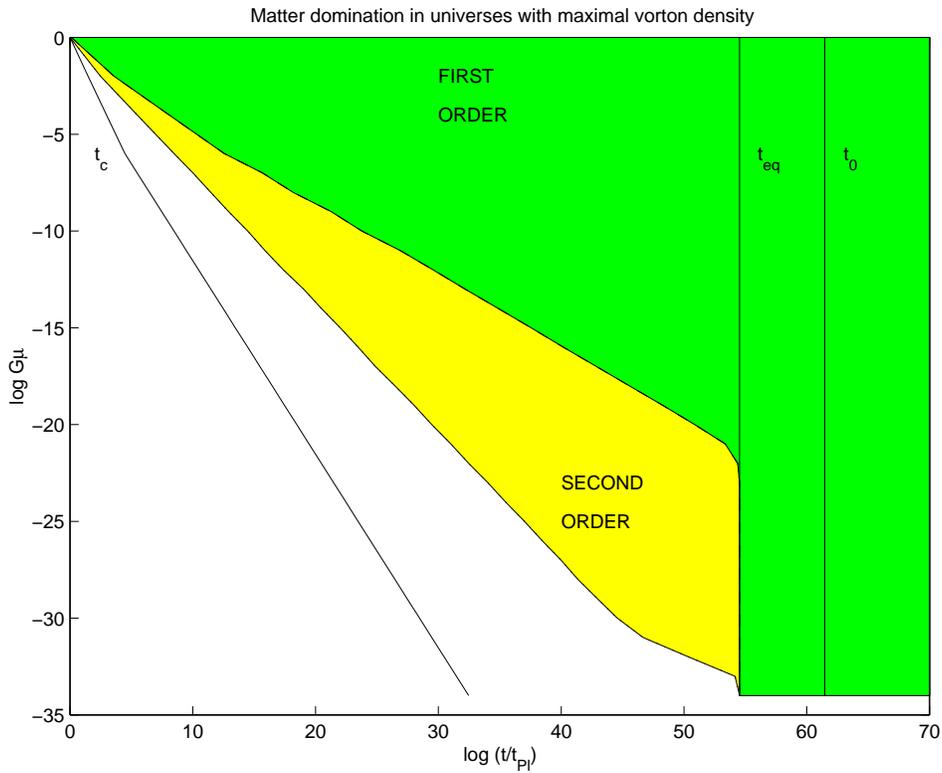}}
\vskip.3in}
\caption{Cosmological properties of {\em maximal} vorton-producing string networks.
This shows, for all allowed values $G\mu$, the epoch at which each network
would form, in orders of magnitude after the Planck time ($t_c$ line), and the epoch of onset of matter-domination in such a universe for initial conditions typical of first (dark grey region) and second (extended light gray) order phase transitions. Also shown are the epochs of equal matter and radiation densities in the 'standard' cosmological scenario and the present epoch ($t{eq}$ and $t_0$ lines, respectively). Note that this plot is an {\em incorrect} overestimate of vorton production. (See text for explanation.)}
\label{figure1}
\end{figure}

Before discussing what we believe is the right result, however, let us
discuss a naive `maximal' argument that has appeared a number of times in the literature. We start by analysing the {\em hypothetical} case where all
loops produced by the string network become vortons. The results in such a case
are summarised in fig. \ref{figure1}. This shows, for different values of the string mass parameter $G\mu$ (ranging from Planck-scale strings with $G\mu\sim1$ to electroweak-scale strings with $G\mu=10^{-34}$) the epoch at which each network would form, in orders of magnitude after the Planck time (the dotted line marked $t_c$), and the epoch at which a universe with such a string network would become matter-dominated for initial conditions typical of first (dashed
curve) and second (solid curve) order phase transitions. Also shown are the
epochs of equal matter and radiation densities in the `standard' cosmological scenario and the present epoch (dotted lines marked $t{eq}$ and $t_0$, respectively). Note the strong dependence on initial conditions, as claimed.
If the string-forming and superconducting phase transitions are of second order, then all vorton-forming strings above the electroweak scale would be
excluded, as vorton production would make the universe matter-dominated too early to be consistent with observations. On the other hand, if the phase transitions were of strongly first order, then only strings heavier than $G\mu=10^{-20}$ (that is, those formed above $T_c\sim10^9\,GeV$) would be excluded.

However, we strongly emphasise that fig. \ref{figure1} is {\em wrong} since,
as we claimed above (and have shown in \cite{vornew}) this `perfect efficiency'
is unrealistic. There is a considerable region of the parameter space of possible loop initial conditions (which is largest for GUT-scale strings and whose importance in the total available phase space decreases with the string mass parameter $G\mu$) for which loops are not prevented from becoming relativistic and hence no vortons form. On the other hand, there is a minimum size for loops, below which they will simply disintegrate. The reason for us to present fig. \ref{figure1} is to emphasise that in this type of cosmological problem one can easily be led to wrong results by choosing assumptions that are too optimistic (as was done by a number of authors in the past \cite{vor,vor2}).

Having made this point (we hope) perfectly clear, let us now discuss what we
believe is the correct answer to the `vorton problem', or at least the best
one that can be presently given, pending a detailed study of the question of
quantum mechanical stability of these objects. As we already mentioned, the
problem with fig. \ref{figure1} is at the high-mass end: in this region, the
friction-dominated epoch is very short and there is not time for loops to build
up high enough currents to prevent them from becoming relativistic. Hence in
this region we do not expect any vortons to form, so that no cosmological
constraints on these superconducting models can be set. As was already pointed
out in \cite{vornew}, this region includes GUT-scale string.

When vortons start
to be produced, however, their high mass means that even a very small number of
them is enough to start matter domination. Hence there is an intermediate $G\mu$
range where we can rule out Witten-type superconducting models. It should be
noted, however, that the excluded region is strongly dependent on initial
conditions---in particular, on the order of the string-forming phase transition. For a first-order phase transition, the exclusion region
corresponds to energy scales $T_c\sim10^9-10^{12}\,GeV$, while for a
second-order one it is much larger, $T_c\sim10^5-10^{14}\,GeV$. Finally, for lower-mass superconducting strings, vortons can be a source of dark matter. Still, note again the importance of the initial conditions. At the electroweak scale (where essentially all but the smallest loops become vortons), the present density of the universe in the form of vortons can be up to
$\Omega_v\sim0.06$ if the phase transitions were of second order, but it will
only be $\Omega_v\sim10^{-16}$ in the most unfavourable case of first-order
phase transitions.  Finally, we point out that a superconducting cosmic string
network formed at an energy scale $T_c\sim10^4-10^8\,GeV$ can (depending on
initial conditions), provide a rather simple and elegant solution to the dark
matter problem of the universe. This being so, the recently popular low-energy
super-symmetry-breaking models which admit superconducting cosmic strings
(see for example \cite{riotto}) might well prove useful.

Finally, we provide a simple fitting formula which approximately reproduces our results (see fig. \ref{figure4}). Choose a string-forming energy scale $T_c$, or equivalently
\begin{equation}
G\mu\sim\left(\frac{T_c}{m_{Pl}}\right)^2\, .\label{defgm}
\end{equation}
Then the initial correlation length of the string network must obey (see \cite{ms})
\begin{equation}
\ell_{\rm f}\lsim L_c \lsim \ell_H\, ,\label{inlenght}
\end{equation}
where $\ell_H=t$ is the horizon length and $\ell_{\rm f}\sim(G\mu)^{1/2}$ is the friction length. One expects strings formed in strongly first order phase transitions to be close to the above upper limit, while those formed in second order transitions should be close to the lower one.  We therefore introduce a parameter $s$ characterising the `strength' of the phase transition and defined as follows
\begin{equation}
s=1-\frac{\log(L_c/t_c)}{\log(\ell_{\rm f}/t_c)}\, .\label{defess}
\end{equation}
This will therefore vary between zero (second order transition, $L_c\sim (G\mu)^{1/2}t_c$) and unity (strong first order transition, $L_c\sim t_c$). Given these, our results indicate that no vortons form if $\log(G\mu)\gsim  -4s-10$. On the other hand, the region
\begin{equation}
-13.3s^2+21.3s-28\lsim \log(G\mu)\lsim  -4s-10\, \label{exclregion}
\end{equation}
is excluded, since vortons would become the dominant component of the universe, leaving a fraction of the critical density in baryons that is incompatible with nucleosynthesis.
Finally, for low-mass string networks, $\log(G\mu)\lsim -13.3s^2+21.3s-28$,
our results for the present density in vortons can be reproduced by the following expression
\begin{equation}
\Omega_V(t_0)\simeq 10^{a_1 (\log G\mu)^2+a_2 \log G\mu +a_3}\, , \label{fformula}
\end{equation}
where
\begin{equation}
a_1=0.011s-0.029\, ,\label{defa1}
\end{equation}
\begin{equation}
a_2=1.81s-1.62\, ,\label{defa2}
\end{equation}
\begin{equation}
a_3=33.63s-22.29\, .\label{defa3}
\end{equation}
However, a word of caution is needed here. This fit is accurate in the limits $s=0$ and $s=1$, but less accurate in intermediate cases. In this case, a  `scan' of the parameter space of possible initial conditions shows that the above expression is accurate at the order of magnitude level, but not more than that. Hence it should be used with some care, in particular when one is interested in cases that lie close to the `border' between the excluded and dark matter producing regions (in which case a detailed calculation should be required). Finally, we emphasise that our results apply only to Witten-type superconducting string models. As was pointed out elsewhere \cite{vornew}, strong model dependence is expected.

\section{Conclusions}
\label{v-ccl}
In this letter we have used a recently developed \cite{vornew} semi-analytic model for vorton formation to study the mechanism of chiral vorton production
in Witten-type superconducting string models of all relevant masses. We have
numerically confirmed the claim \cite{vornew} that in addition to a low-energy
regime where vortons can be a source of dark matter and an intermediate regime
where vorton-producing networks are observationally excluded there is also a
high-energy regime where no vortons are produced.

We have emphasised that there
is a crucial dependence of the results on the initial conditions of the
formation of the string networks. Nevertheless, we are able to rule out any vorton-forming cosmic strings in the range  $10^{-20}\lsim G\mu
\lsim 10^{-14}$ (that is $10^{9}\hbox{GeV}\lsim T_{\rm c}
\lsim 10^{12}\hbox{GeV}$), irrespective of initial conditions. For the particular case of a second-order string transition (producing a very high-density string network), vorton production will be cosmologically disastrous for the range $10^{-28}\lsim G\mu
\lsim 10^{-10}$ (or $10^{5}\hbox{GeV}\lsim T_{\rm c}
\lsim 10^{14}\hbox{GeV}$). These bounds assume that the energy scales of the string and current-forming phase transitions are not too widely separated.

On the other hand, cosmic string networks formed slightly below the lower limit of the excluded energy region can provide a significant contribution to the dark matter of the universe. We therefore put forward the claim that cosmic strings produced in low-energy super-symmetry-breaking models are good candidates for solving the dark matter problem of the universe. We have also provided a simple fitting formula which approximately summarises our results.

To conclude, let us emphasise that there is a crucial aspect of
the vorton phenomenology that still remains to be considered---namely, that
of their quantum mechanical stability. Obviously, this is a very difficult and
delicate problem, which is the reason why it has been largely untouched for
nearly a decade since the first references to vortons appeared \cite{vor}.
However, it is clear that any further attempt of a more detailed study of
vorton cosmology must necessarily start in that direction, and we hope to
concentrate our own efforts on this important task.

\acknowledgments
C.M.\ is funded by FCT
(Portugal) under `Programa PRAXIS XXI' (grant no.
PRAXIS XXI/BD/11769/97). E.P.S.\ is funded by PPARC and
we both acknowledge the support of PPARC and the
EPSRC, in particular the Cambridge Relativity rolling
grant (GR/H71550) and a Computational Science
Initiative grant (GR/H67652).

\begin{figure}
\vbox{\centerline{
\epsfxsize=0.7\hsize\epsfbox{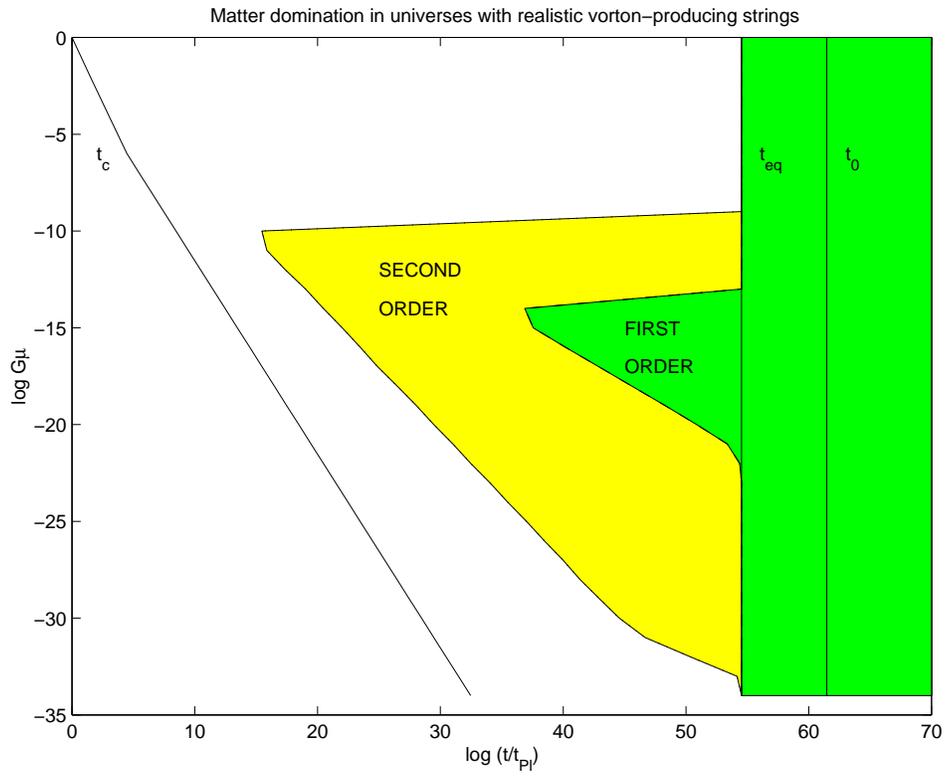}}
\vskip.3in}
\caption{Same as fig.1, but for {\em realistic} Witten-type superconducting strings
({\em i.e.}, this is the {\em correct} plot). The string network forms at $t_c$, and the onset of matter domination is brought forward to unacceptably early times for an intermediate $G\mu$ range, which depends on the phase transition order.}
\label{figure2}
\end{figure}

\vfill\eject

\begin{figure}
\vbox{\centerline{
\epsfxsize=0.7\hsize\epsfbox{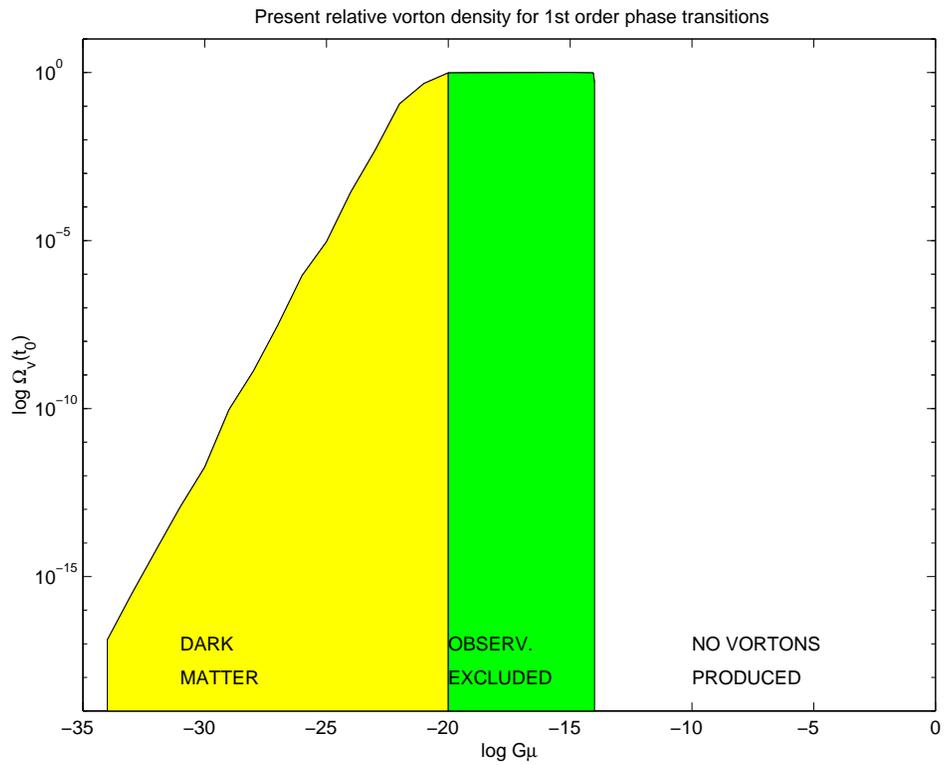}}
\vskip.3in}
\vbox{\centerline{
\epsfxsize=0.7\hsize\epsfbox{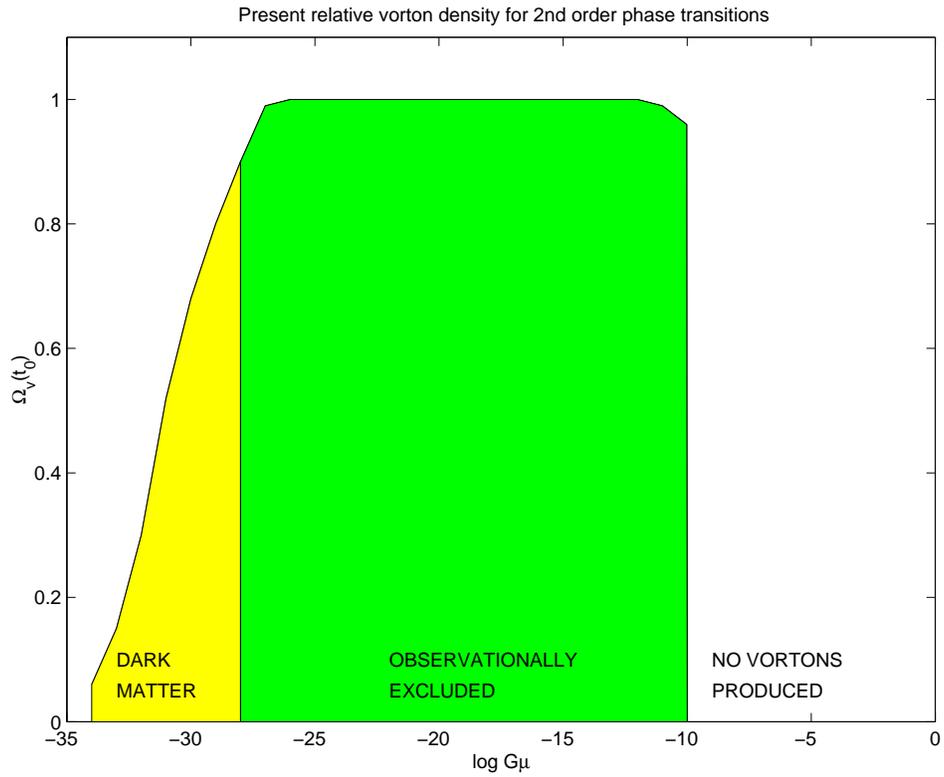}}
\vskip.3in}
\caption{The present contributions of chiral vortons to the density of the universe,
for Witten-type superconducting strings produced at first (top) and second 
(bottom) order phase transitions. Note that in  each case there are three different regions
depending on the string mass (see text).}
\label{figure3}
\end{figure}

\vfill\eject

\begin{figure}
\vbox{\centerline{
\epsfxsize=0.7\hsize\epsfbox{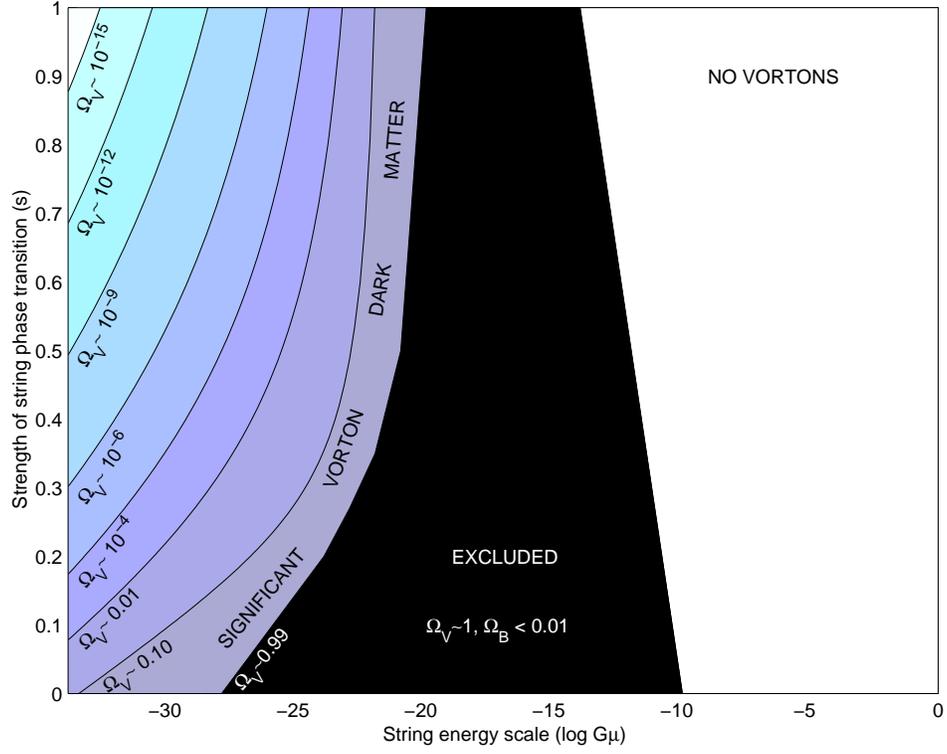}}
\vskip.3in}
\caption{The expected present vorton density, according to the fitting formula (\protect\ref{fformula}), as a function of the string energy scale and the strength of the phase transition that produced them (see text). Several relevant contours are marked.}
\label{figure4}
\end{figure}

\end{document}